\documentclass[a4paper,12pt,reqno]{amsart}
\newtheorem{proposition}{Proposition}
\newtheorem{lemma}{Lemma}

\def\la{\langle}
\def\ra{\rangle}

\def\P{{\mathcal P}}

\def\ds{\displaystyle}

\def\proof{\noindent{\bf Proof.\ }}
\def\quadratino{
\hfill\vbox{\hrule\hbox{\vrule\vbox to 7 pt {\vfill\hbox to
7 pt {\hfill\hfill}\vfill}\vrule}\hrule}\par}

\def\Na{{\mathbb N}}

\def\Re{{\mathbb R}}

\def\sleq{\leq\kern-4pt\cdot}
\def\sgeq{\null\kern+1pt\cdot\kern-4.5pt\geq}

\def\be{\begin{equation}}
\def\ee{\end{equation}}

\newcommand{\ps} {\mathcal{M}}
\newcommand{\ma} {T}
\newcommand{\ts} {\mathrm{T}}

\begin{document}
\baselineskip=19pt

\title[Localization in infinite billiards]{Localization in infinite
billiards: a comparison between quantum and classical ergodicity}
\author{Sandro Graffi $\qquad$ Marco Lenci}
\address{Dipartimento di Matematica, Universit\`a di Bologna  
\newline
\hphantom{vi} Piazza di Porta S. Donato 5, 40127 Bologna, Italy.} 
\email{graffi@dm.unibo.it} 
\address{Department of Mathematical Sciences, Stevens Institute of 
Technology \newline 
\hphantom{vi} Hoboken, N.J. 07030, U.S.A.}
\email{mlenci@math.stevens.edu}
\date{} 

\begin{abstract}
Consider the non-compact billiard in the first quandrant bounded by
the positive $x$-semiaxis, the positive $y$-semiaxis and the graph of
$f(x) = (x+1)^{-\alpha}$, $\alpha \in (1,2]$. Although the Schnirelman
Theorem holds, the quantum average of the position $x$ is finite on
any eigenstate, while classical ergodicity entails that the classical
time average of $x$ is unbounded.
\end{abstract}

\maketitle
\section{Introduction and statement of the result}
\setcounter{equation}{0}%
\setcounter{theorem}{0}%
\setcounter{proposition}{0}%
\setcounter{lemma}{0}%
\setcounter{corollary}{0}%
\setcounter{definition}{0}%
\setcounter{remark}{0}%
\setcounter{example}{0}%

\noindent
The purpose of this note is to exhibit a simple example of a chaotic
system in which the long time classical and quantum evolutions can be
proved to be qualitatively different even though the Schnirelman
Theorem (sometimes called quantum ergodicity: see \cite{Sc} and also
\cite{CdV}, \cite{DGI}, \cite{HMR}, \cite{Ze}, \cite{ZZ}) holds.

Let $Q$ be the planar domain bounded by the positive $x$-semiaxis, the
positive $y$-semiaxis and the graph of $f(x) = (x+1)^{-\alpha}$,
$\alpha \in (1,2]$. Notice that $\mathrm{Area}(Q)<+\infty$. Let
$\phi^t: Q \times S^1 \longrightarrow Q \times S^1$ be the dynamical
flow corresponding to the billiard motion in $Q$. This is the
Hamiltonian flow of a particle of energy 1 that moves freely in $Q$
and performs totally elastic collisions at the boundary. By \cite{Le2}
$\phi^t$ is ergodic w.r.t.\ the normalized Liouville measure $d\nu :=
(2\pi \, \mathrm{Area}(Q))^{-1} \, dx\, dy\, d\theta$. (Ergodicity
there is proved for some Poincar\'e sections and then extended to the
billiard flow; see also \cite{Le3} for a general class of ergodic
billiards of the like, of finite or infinite area, including some that
qualify as further examples for the present paper.)  Moreover, as we
show in the Appendix below, the Lyapunov exponent of the flow is
positive. Therefore this system is chaotic according to general
consensus.

In quantum mechanics, the corresponding system is defined by the
Schr\"odinger operator
\begin{equation}
  H := -\hbar^2\Delta_D, \qquad D(H)=H^2(Q)\cap H_0^1(Q) \subset L^2(Q),
\end{equation} 
where $\Delta_D$ is the Laplacian with Dirichlet boundary conditions
at $\partial Q$. It is an old result of Rellich that $\Delta_D$ is
self-adjoint with compact resolvent. Hence $\mathrm{Spec}(H)$ is
discrete. (More recent results on $\mathrm{Spec}(\Delta_D)$ include
\cite{Ta}, \cite{Ro}, \cite{Si}, \cite{vdB}, \cite{Da}).  We denote 
$E_j(\hbar) = \hbar^2 \ell_j$, $j \in \Na$, the eigenvalue of $H$
corresponding to the normalized eigenfunction $\psi_j$. The order is
such that $j \mapsto \ell_j$ is non-decreasing and
$\ds \{\psi_j\}_{j\in\Na}$ is a complete orthonormal system.

The classical time average of the coordinate $x$ is
\begin{equation}
\label{mediac}
  \overline{X} (x,y,\theta) := \lim_{T \to +\infty} \frac1T 
  \int_0^T X \circ \phi^t (x,y,\theta) \, dt,
\end{equation}
where $X(x,y,\theta) := x$. In principle one does not know whether
this limit exists for almost all $(x,y,\theta)$, as $X \not\in L^1(Q
\times S^1)$ and Birkhoff's Theorem does not apply. However,
Proposition \ref{P} below will override this question.

Tolerating a little abuse of notation, the quantum expectation of the
position operator $X$ on any $u \in L^2(Q)$ is instead
\begin{equation}
\label{mediaq}
  \langle X \rangle (u) := \la u, Xu \ra = \int_Q x |u(x,y)|^2 dxdy,
\end{equation}

The comparison result we want to prove can be formulated as follows.

\begin{proposition}
\label{P}
\par\noindent
\begin{itemize}
\item[(a)] The classical average is infinite, i.e., for $\nu$-a.e.\
initial condition $(x,y,\theta)$:
\begin{equation}
  \overline{X} (x,y,\theta) = +\infty.
\end{equation}
\item[(b)] The eigenfunctions of $\Delta_D$ are
\emph{super-exponentially localized} in the following sense: Denote
\begin{equation}
  \xi_j(x) := \int_0^{f(x)} |\psi_j(x,y)|^2 dy,
  \label{xi}
\end{equation}
the probability density of the position in the $x$-direction. Then one
has:
\begin{equation}
  \forall \gamma>0, \quad \xi_j(x) = o (e^{-\gamma x}),
  \label{super-exp}
\end{equation}
as $x \to +\infty$. Therefore, in particular, the quantum expection of
$X$ on every eigenstate is finite:
\begin{equation}
  \langle X \rangle (\psi_j) := \int_Q x |\psi_j (x,y)|^2 dxdy =
  \int_0^{+\infty} x \, \xi_j(x) dx < +\infty.
  \label{qloc}
\end{equation}
\item[(c)] The Schnirelman Theorem holds. As a corollary, there is a
density-1 sequence $\{ j_n \}$, i.e., a sequence with the property  
\begin{equation}
  \lim_{k \to +\infty} \frac{\# \{ j_n \le k \}} k = 1,
\end{equation}
such that
\begin{equation}
  \lim_{n \to +\infty} \langle X \rangle (\psi_{j_n}) = +\infty.
  \label{schn-x}
\end{equation}
\end{itemize} 
\end{proposition}

\newpage

{\bf Remarks}
\par\noindent
\begin{enumerate}

\item Proposition \ref{P},(b) implies that $\langle F(X) \rangle (u) <
+\infty$ for any (normalized) state $\ds u := \sum_{j=1}^\infty a_j
\psi_j$ with $j \mapsto |a_j|$ decaying sufficiently fast. Here $F \in
L^1_{loc}(\Re)$ is any $x$-dependent observable exponentially bounded
as $x\to +\infty$.  In fact, a Cauchy-Schwartz inequality shows that
\begin{eqnarray}
   | \la u, F(X)u \ra | &\le& \| F(X)u \| \label{loc} \\
   &\le& \sum_{j=1}^\infty |a_j| \left( \int_0^{+\infty} |F(x)|^2 
   \, \xi_j(x) dx \right)^{1/2} < +\infty. \nonumber
\end{eqnarray}
If we replace $F(X)$ by the corresponding Heisenberg observable $\ds
F(X)(t) := e^{iH t/\hbar} F(X) e^{-iH t/\hbar}$, clearly the bound
(\ref{loc}) holds uniformly in $t$:
\begin{eqnarray}
  | \la u, F(X)(t) u \ra | &=& | \la e^{-iH t/\hbar} u, F(X) e^{-iH
  t/\hbar}u \ra | \\ 
  &\le& \sum_{j=1}^\infty |a_j| \left( \int_0^{+\infty} |F(x)|^2 \,
  \xi_j(x) dx \right)^{1/2} < +\infty. \nonumber
\end{eqnarray}
This last estimate shows that the quantum evolution is also
super-exponen\-tially localized.

\item Therefore, for the same $u$ as above, one also obtains
convergence of the time average of the Heisenberg observable $X(t)$:
\begin{eqnarray}
  \overline{\langle X \rangle} (u) &:=& \lim_{T\to\infty} \frac1T
  \int_0^T \la u, e^{iHt/\hbar} X e^{-iHt/\hbar} u \ra \nonumber \\
  &=& \sum_{{j,k \in \Na} \atop {\ell_j = \ell_k}} a_j^{*} a_k \la
  \psi_j, X \psi_k \ra < +\infty.
\end{eqnarray}
This Heisenberg time average $\overline{\langle X \rangle}$ is the
quantity to be directly compared with the classical time average
$\overline{X}$.

\item Assertion (b) shows that quantum mechanics localizes the
unbounded classical chaotic motion as soon as $\hbar >0$. This
phenomenon is an example of quantum suppression of classical
chaos. Its physical origin is clear: when the quantum particle visits
the deepest recesses of the cusp, its $y$-position is very well
determined. Therefore, by the Uncertainty Principle, its $y$-velocity
must be extremely spread out, and this cannot occur at finite
energies. (This crucially depends on the Dirichlet boundary
conditions: for the Neumann Laplacian the situation is quite
different \cite{DS}.) One might think of this phenomenon as the
opposite of the tunnel effect: the classical particle is much more
likely to ``penetrate'' the cusp than the quantum one.

\item The classical limit is naturally defined as the joint limit
$j \to +\infty$, $\hbar \to 0$, such that $E_j(\hbar) = 1$ (1 being
the fixed value of the energy). Since the quantities at hand here do not
depend on $\hbar$, the second limit can be forgotten. Obviously,
(\ref{super-exp})-(\ref{qloc}) are not uniform as $j \to +\infty$. 

\item In part (c) the Schnirelman Theorem is formulated for the first
time for systems whose classical energy surface is not compact
(although one does not really need more than \cite{ZZ} to prove
it in our case---see below). As is well known, the Schnirelman
Theorem is a purely asymptotic statement. The present example shows
moreover that this asymptotic statement cannot exclude that
quantization can turn a classical behavior at infinity into a
behavior of a completely different nature. Hence, in general, it might
appear physically misleading to call quantum ergodicity the simple
validity of the Schnirelman Theorem.
\end{enumerate}

\section{Proof of the Proposition}
\setcounter{equation}{0}%
\setcounter{theorem}{0}%
\setcounter{proposition}{0}%
\setcounter{lemma}{0}%
\setcounter{corollary}{0}%
\setcounter{definition}{0}%
\setcounter{remark}{0}%
\setcounter{example}{0}%

\noindent
(a) Since clearly
\begin{equation}
  \int_{Q \times S^1} X\, d\nu = \frac1 {\mathrm{Area}(Q)}
  \int_0^{+\infty} x f(x) dx = +\infty.
\end{equation}
the assertion follows follows directly from classical ergodicity
\cite{Le2} and the following easy lemma.

\begin{lemma}
Let $(\P,\nu)$ be a probability space and $\phi^t$ a flow on $\P$ that
preserves the measure $\nu$. If the measurable function $g$ is bounded
below and $(\P, \phi^t, \nu)$ is ergodic then, for $\nu$-a.e.\
$z\in\P$,
\begin{equation}
  \lim_{T \to +\infty} \frac1T \int_0^T g \circ \phi^t (z) \, dt =
  \int_\P g \, d\nu,
\end{equation}
whether $g$ is integrable or not.
\label{tech-lem}
\end{lemma}

\proof If $g$ is integrable there is nothing to prove by
ergodicity. Otherwise $\int_\P g \, d\nu = +\infty$, because $g$ is
bounded below. In this case, for $m \in \Na$, set $g_m := i_m \circ
g$, with $i_m(x) := x$ for $x \le m$ and $i_m(x) := m$ for $x > m$. By
the Birkhoff theorem, there is a full-measure set $B_m$ over which the time
average of $g_m$ exists and is equal to its phase average.  Therefore,
for $\ds z \in B:= \bigcap_{m \in \Na} B_m$ (still a full-measure
set),
\begin{eqnarray}
  \liminf_{T \to +\infty} \frac1T \int_0^T g \circ \phi^t (z)
  \, dt, &\ge& \liminf_{T \to +\infty} \frac1T \int_0^T g_m \circ 
  \phi^t (z) \, dt \nonumber \\
  &=& \int_\P g_m \, d\nu.
\end{eqnarray}
By monotonic convergence, the sup of the r.h.s.\ in $m$ is $+\infty$,
while the l.h.s.\ does not depend on $m$. This proves at once that the
limit in $T$ exists and is $+\infty$.
\quadratino
\medskip\noindent 
(b) For the proof of this part, let us drop the subscript $j$ from all
notation. Hence $\psi$ is continuous on $Q$, infinitely smooth in its
interior, and such that
\begin{equation}
  - (\partial_x^2 + \partial_y^2) \psi(x,y) = \ell \psi(x,y),
  \label{eigen}
\end{equation}
with $\psi |_{\partial Q} = 0$. Without loss of generality $\psi$ is
real.  Recalling definition (\ref{xi}) one easily checks, by repeated
differentiation inside the integral, that
\begin{equation}
  \xi''(x) = \int_0^{f(x)} \partial_x^2 \psi^2 dy \ge 2 \int_0^{f(x)}
  \psi \partial_x^2 \psi dy.
  \label{eq1}
\end{equation}
Now, think of $-\partial_y^2$ as the $1-$dimensional Laplacian on $[0,a]$
with Dirichlet boundary conditions. Then, in the sense of the quadratic
forms, $-\partial_y^2 \ge (\pi/a)^2$. Therefore, multiplying
(\ref{eigen}) by $\psi$, integrating over $y$, and using that lower
bound, we obtain
\begin{eqnarray}
  -\int_0^{f(x)} \psi \partial_x^2 \psi dy &=& \ell \int_0^{f(x)}
  \psi^2 dy + \int_0^{f(x)} \psi \partial_y^2 \psi \nonumber \\ 
  &\le& \left[ \ell - \left( \frac{\pi}{f(x)} \right)^2 \right]
  \int_0^{f(x)} \psi^2 dy.
\end{eqnarray}
Plugging into (\ref{eq1}),
\begin{equation}
  \xi''(x) \ge 2 \left[ \left( \frac{\pi}{f(x)} \right)^2 - \ell \right]
  \xi(x) \ge \gamma^2 \xi(x),
  \label{eq2}
\end{equation}
for any $\gamma>0$, provided $x$ is large enough depending on $\gamma$.
This means that either $\xi(x) \ge C e^{\gamma x}$ or $\xi(x) \le C
e^{-\gamma x}$. Since $\psi$ is an eigenfunction, the first possibility
cannot occur. (\ref{super-exp}) is thus proved.

\medskip\noindent 
(c) The version of the Schnirelman Theorem that can be verified in our
case is the following: For any pseudodifferential operator $A$ of
order 0, compactly supported in the position variables, there exists a
density-1 sequence $\{ j_n \}$ such that
\begin{equation}
  \lim_{n\to +\infty} \la \psi_{j_n}, A \, \psi_{j_n} \ra = \int_{Q
  \times S^1} a \, d\nu,
  \label{schn}
\end{equation}
where $a$ is the principal symbol of $A$. The proof is a
strightforward check that all arguments of \cite{ZZ} valid for compact
billiards hold true in this case too. We limit ourselves to remark
that the crucial fact is that $\mathrm{Area}(Q) < +\infty$, and in
particular that the system is recurrent (see \cite{Le1} and references
therein for better results in this direction).

To prove (\ref{schn-x}) we use an argument that is somewhat similar to
Lemma \ref{tech-lem}. Let $\{ X_m \}_{m\in\Na}$ be an increasing
sequence of functions on $Q \times S^1$ such that: $X_m$ is smooth and
depends only on $x$, $\mathrm{supp}(X_m) \subseteq \{x \le m\}$, $X_m
\le X$ and, pointwise, $X_m \to X$ for $m \to +\infty$. With the usual
abuse of notation, we denote $X_m$ also the corresponding
multiplication operator on $L^2(Q)$.

For every $m$ there is a density-1 sequence $\{ j_n^{(m)} \}_{n\in\Na}
=: \sigma^{(m)}$ such that (\ref{schn}) holds for $A = X_m$ (thus $a =
X_m$), using that sequence. Now fix $p_1 = 0$ and, for $m \ge 2$,
consider the following recursive definition: Select a sufficiently
large $p_m > p_{m-1}$ so that
\begin{equation}
  \forall j_n^{(m)} \ge p_m, \qquad \left| \la \psi_{j_n^{(m)}}, X_m \,
  \psi_{j_n^{(m)}} \ra - \int_{Q \times S^1} X_m \, d\nu \right| \le
  \frac1m,
  \label{rec1}
\end{equation}
and
\begin{equation}
  \frac{\# \left(\sigma^{(m-1)} \cap [p_{m-1}, p_m) \right) } {p_m -
  p_{m-1}} \ge \frac1m
  \label{rec2}
\end{equation}
(here we have identified $\sigma^{(k)}$ with its image). (\ref{rec2})
can always be verified because $\sigma^{(k)} \cap [p_k, +\infty)$ has
density 1.

Now set $\ds \sigma := \bigcup_{m \in \Na} \left( \sigma^{(m)} \cap
[p_m, p_{m+1}) \right)$. Think of $\sigma$ as a sequence and label its
elements $\{ j_n \}$. By (\ref{rec2}), $\sigma$ has density
1. Moreover, by (\ref{rec1}),
\begin{equation}
  \la \psi_{j_n}, X \, \psi_{j_n} \ra \ge \la \psi_{j_n}, X_m \,
  \psi_{j_n} \ra \ge \int_{Q \times S^1} X_m \, d\nu - \frac1m,
\end{equation}
if $j_n \in [p_m, p_{m+1})$. Therefore the $\liminf$ of the l.h.s.\ in
$n$ corresponds to the $\liminf$ of the r.h.s.\ in $m$: the latter is
infinity by monotonic convergence.

\section{Appendix: Positivity of the Lyapunov exponent} 
\setcounter{equation}{0}%
\setcounter{theorem}{0}%
\setcounter{proposition}{0}%
\setcounter{lemma}{0}%
\setcounter{corollary}{0}%
\setcounter{definition}{0}%
\setcounter{remark}{0}%
\setcounter{example}{0}%

\noindent
In this section we prove that, as claimed in the Introduction, 
$\phi^t$ has a positive Lyapunov exponent $\lambda$ (and thus a 
negative exponent $-\lambda$, as $\phi^t$ is volume-preserving and 
the exponent in the direction of the flow is obviously 0).

For $z := (x,y,\theta) \in Q \times S^1$, denote $E^c_z \subset \ts_z
(Q \times S^1)$ the one-dimensional subspace of the tangent space at
$z$ generated by $\partial_t \phi^t(z)$, i.e., the direction of the
flow in $z$. (The inconsequential complications arising when $(x,y)
\in \partial Q$ are ignored). It is known that $D\phi^t$ preserves the
splitting $E^c \oplus (E^c)^\perp$ (see \cite{SC}---this corresponds
to the fact that the wave-front of a light wave is always orthogonal
to its rays). Therefore the stable and unstable directions at $z$
(denoted $E^s_z$ and $E^u_z$, respectively) are to be looked for
within $(E^c_z)^\perp$. To verify their existence we use the results
of \cite{Le2} about $\ps$, the Poincar\'e section given by all unit
vectors (\emph{line elements}, in the language of \cite{S}) that are
based in $\partial Q$ and point towards the interior of $Q$. We know
that $\ma$, the Poincar\'e map of $\ps$, has stable and unstable
directions, denoted $F^{s(u)}$: we will use these to define $E^{s(u)}$
everywhere (more precisely, in a full-measure subset of $Q \times
S^1$).

Consider a non-singular element $z \in \ps$, that is, a line element
whose future orbit never hits $\partial Q$ tangentially or intersects
the vertex at $(0,1)$. (Singular line elements can be neglected
because they form a measure-zero set in $\ps$, and thus in $Q \times
S^1$, relatively to their respective invariant measures).  Since $z$
is non-singular, $E^c_z \not\subset \ts_z \ps$ and so $\Pi_z : \ts_z
\ps \longrightarrow (E^c_z)^\perp$, the projection in the direction of
$E^c_z$, is non-degenerate. We define $E^{s(u)}_z := \Pi_z
F^{s(u)}_z$. Finally, for $t>0$, $D\phi^t E^{s(u)}_z$ define the
stable and unstable directions at all points in the future orbit of
$z$. It is easy to see that this definition is unambiguous.

For a non-singular $z$ (not necessarily in $\ps$), take $v^u \in
E^u_z$. The sought Lyapunov exponent is
\begin{equation}
  \lambda(z) := \lim_{T\to +\infty} \frac1T \log \frac{ | D\phi^T_z
  (v^u) | } {|v^u|} = \lim_{T\to +\infty} \frac1T \log \left\|
  \left. D\phi^T_z \right|_{E^u_z} \right\|.
  \label{lyap1}
\end{equation}
Let us introduce the following notation: $\tau(z) := \min \left\{ t\ge
0 \,\big|\, \phi^t (z) \in \partial Q \times S^1 \right\}$ is the
(forward) \emph{free path} of the line element $z$, $\tau^{-}(z)$ is
the backward free path, analogously defined, and $z' = z'(z) :=
\phi^{-\tau^{-}(z)} (z)$ is the closest line element in the past orbit
of $z$ that is based in the boundary. Define
\begin{equation}
  g(z) := \frac1 {\tau(z')} \log \left\| \left. D\phi^{\tau(z')+}_{z'}
  \right|_{E^u_{z'}} \right\|,
  \label{def-g}
\end{equation}
where $D\phi^{t+}$ means $\ds \lim_{\varepsilon \to 0^{+}} D\phi^{t +
\varepsilon}$. (This is needed because $\phi^t$ is discontinuous at
$\partial Q \times S^1$.)  By construction, $g$ is constant along the
rectilinear parts of a given orbit. Furthermore, due to the composition
properties of the differential,
\begin{equation}
  \lambda(z) = \lim_{T \to +\infty} \frac1T \int_0^T g \circ \phi^t
  (z) \, dt.
  \label{lyap2}
\end{equation}
In fact, in (\ref{lyap1}) and (\ref{lyap2}), the two functions that
are averaged in $T$ can only differ in the first and last segment of
of $\left\{ \phi^t(z) \right\}_{t=0}^T$. This difference gets washed
away as $T \to +\infty$.

We claim that $g > 0$. In order to see this, let us go back to the
Poincar\'e map $\ma$. For $z \in \ps$, we have seen that
\begin{equation}
  \left. D\phi^{\tau(z)+}_z \right|_{(E^c_z)^\perp} = \Pi_{\ma(z)} \,
  D\ma_z \, \Pi^{-1}_z .
  \label{phi-ma}
\end{equation}
But it is a fact that $\left. D\ma \right|_{F^u}$ is strictly
expanding w.r.t.\ to a certain metric in $\ts \ps$. This metric is
called in \cite{Le2} the \emph{increasing metric} and is defined as
$|v|_\mathrm{inc} := | \Pi v|$. This fact and (\ref{phi-ma}) show that
$\left. D\phi^{\tau +} \right|_{E^u}$ is strictly expanding which, in
view of (\ref{def-g}), proves the claim.

Finally we apply Lemma \ref{tech-lem} to conclude that, for
$\nu$-a.e.\ $z \in Q \times S^1$,
\begin{equation}
  \lambda(z) = \int_{Q \times S^1} g \, d\nu > 0.
\end{equation}

\vskip 1.0cm\noindent
{\bf Acknowledgments.} We thank N.~Burq, G.~Forni, J.~Wunsch and
M.~Zworski for some very useful discussions. 

\end{document}